\documentclass[12pt,titlepage]{article}
\usepackage[USenglish]{babel}
\usepackage{sectsty}
\usepackage{amsfonts}
\usepackage{amsmath}
\usepackage{amssymb}
\usepackage{geometry}
\usepackage{apacite}
\usepackage{setspace}
\usepackage{apalike}
\usepackage{float}
\usepackage{portland}
\usepackage{scalefnt}

\input{tcilatex}
\begin{document}

\noindent George Karabatsos* and Stephen G. Walker\medskip \smallskip 
\newline
\begingroup%
\scalefont{1.4}%
\textbf{A Bayesian Nonparametric Causal Model for Regression Discontinuity
Designs}\footnote{%
This research is supported by NSF grant SES-1156372. Thanks to Phillip Dawid
and Peter M\"{u}ller for helpful comments. The results of this paper were
presented in a University of Texas statistics seminar during Fall of 2013;
and in sessions on causal analysis for the ISBA\ conference during Summer of
2014 in Cancun; for the Society of Research on Educational Effectiveness
(SREE) Spring 2014 conference in Washington, D.C.; and for JSM 2013 at
Montreal.}%
\endgroup%

\begin{center}
February 5, 2015
\end{center}

\bigskip

\noindent \noindent \textbf{Abstract:}\ \ For non-randomized studies, the
regression discontinuity design (RDD) can be used to identify and estimate
causal effects from a "locally-randomized" subgroup of subjects, under
relatively mild conditions. However, current models focus causal inferences
on the impact of the treatment (versus non-treatment)\ variable on the mean
of the dependent variable, via linear regression. For RDDs, we propose a
flexible Bayesian nonparametric regression model that can provide accurate
estimates of causal effects, in terms of the predictive mean, variance,
quantile, probability density, distribution function, or any other chosen
function of the outcome variable. We illustrate the model through the
analysis of two real educational data sets, involving (resp.)\ a sharp RDD
and a fuzzy RDD.

\bigskip

\newline
\newline
--------------\newline
{\small *Corresponding author: George Karabatsos, \noindent 1040 W. Harrison
St. (MC\ 147), University of Illinois-Chicago, IL 60607, USA. E-mail:
gkarabatsos1@gmail.com}\newline
{\small Stephen G.\ Walker, The University of Texas at Austin, 1 University
Station, B6500, USA. E-mail:\ s.g.walker@math.utexas.edu }

\section{Introduction}

A basic objective in scientific research is to infer causal effects from
data. Randomized studies are the gold standard of causal inference. In an
ideal randomized study, the investigator randomly assigns each subject into
one of the treatment conditions, with equal probability, and each subject
complies with her/his treatment assignment. Then, treatment subjects are the
same as non-treatment subjects, in terms of the distribution of all observed
and unobserved pretreatment covariates, aside from sampling error (e.g.,
Rubin, 2008\nocite{Rubin08}); and the outcome variable is independent of the
chosen treatment intervention, conditionally on the treatment variable
(Dawid, 2002\nocite{Dawid02}). Then the causal effect is given by a
comparison of the outcome variable under the treatment intervention, against
the outcome variable under the non-treatment intervention.

Often, it is necessary to estimate causal effects from a non-randomized,
observational study, because a randomized study can be infeasible due to
financial, ethical, or time constraints (Rubin, 2008\nocite{Rubin08}).
However, causal inference from a non-randomized study is more challenging.
This is because without randomization, treated and non-treated subjects
differ almost-surely in terms of the pretreatment covariates.

The regression discontinuity design (RDD)\ (Thistlewaite \&\ Campbell, 1960%
\nocite{ThistlewaiteCampbell60}; Cook, 2008\nocite{Cook08}) is a type of
non-randomized design where a continuous-valued assignment variable (Lee \&\
Lemieux, 2010\nocite{LeeLemieux10}) assigns each subject to the treatment
(non-treatment, resp.)\ condition, whenever her/his observed value of the
assignment variable equals or exceeds (resp. is less than) a fixed cutoff
value. Under relatively mild conditions, notably when subjects have
imperfect control of the assignment variable, the RDD provides a
"locally-randomized experiment." Then treatments are "as good as randomly
assigned" for the subgroup of subjects with assignment variable values near
the cutoff (Lee, 2008\nocite{Lee08}), making the causal effect identifiable
for that subgroup. As proven in Goldberger (2008\nocite{Goldberger08}), the
RDD can empirically produce causal effect estimates that are similar to
those estimates of a standard randomized study (Aiken, et al. 1998\nocite%
{AikenWestSchwalmCarrollHsiung98}; Buddelmeyer \&\ Skoufias, 2004\nocite%
{BuddelmeyerSkoufias04}; Black et al. 2005\nocite{BlackGaldoSmith05};
Schochet, 2009\nocite{Schochet09}; Berk, et al. 2010\nocite%
{BerkBarnesAhlmanKurtz10}; Shadish et al., 2011\nocite%
{ShadishGalindoWongSteinerCook11}).

The RDD has existed for over 50 years, with little initial interest (Cook,
2008\nocite{Cook08}). However, since 1997, more than 74 RDD-based empirical
studies have emerged from these fields (Lee \&\ Lemieux, 2010\nocite%
{LeeLemieux10}; Bloom, 2012\nocite{Bloom12}; Wong et al. 2013\nocite%
{WongSteinerCook13}), for at least three reasons (vander Klaaw, 2008\nocite%
{vanderKlaaw08}; Lee \&\ Lemieux, 2010\nocite{LeeLemieux10}). First, many
non-randomized studies employ treatment assignment rules that can be easily
conceptualized as RDDs. Second, the empirical results of RDDs are intuitive
and can be easily conveyed graphically, say, by a plot of the outcomes
against the assignment variable. Third, the identification of causal effects
in a RDD requires weaker and hence more credible assumptions, compared to
the stronger assumptions that are required by other popular causal models,
mentioned below. This gives the researcher the flexibility to choose from a
range of causal estimation methods.

The other popular causal models for non-randomized studies assume a
"potential outcomes" (counterfactual)\ framework of causal inference (e.g.,
Rubin, 1974\nocite{Rubin74}, 1978b\nocite{Rubin78b}). This is typically done
assuming using notation that is simplified by the Stable Unit Treatment
Value Assumption (SUTVA), which implies no interference between subjects and
no versions of treatments (Rubin, 1990\nocite{Rubin90b}). The popular models
make further assumptions of unconfoundedness (i.e., treatment and
non-treatment outcomes are independent of treatment assignments,
conditionally on all pretreatment covariates) and overlap (i.e., there is a
chance to receive either the treatment or the non-treatment, conditionally
on any value of the pretreatment covariates)\ (Imbens, 2004\nocite{Imbens04}%
). These models are defined by a regression of the outcome variable, on
variables of treatment receipt and observed pretreatment characteristics,
and/or involves matching/weighting subjects on the observed pretreatment
variables and/or on propensity scores (e.g., Imbens, 2004\nocite{Imbens04}).
The regression may also be on a hypothesized set of unobserved pretreatment
covariates, in order to study the sensitivity of causal effect estimates
over varying degrees of hidden bias (e.g., Rosenbaum \&\ Rubin, 1983\nocite%
{RosenbaumRubin83b}), i.e., over changes in the distribution of these
covariates. However, it may be argued that for typical non-randomized
studies, unconfoundedness and overlap are not very credible assumptions
(e.g., Imbens, 2004\nocite{Imbens04}; Lee, 2008\nocite{Lee08}). Even SUTVA
is questionable.

For RDDs, the mainstream causal models are linear, polynomial, or
local-linear models that employ a regression of the outcome variable on the
assignment variable. Such models aim to provide causal inferences in terms
of mean comparisons of treatment outcomes and non-treatment outcomes, and to
provide sufficiently flexible modeling of the regression function (Imbens,
2004\nocite{Imbens04}; Lee \&\ Lemieux, 2010\nocite{LeeLemieux10}), in a
neighborhood around the cutoff. However, in many settings, it may also be of
interest to base causal inferences on comparisons of additional features of
the outcome variable, such as the variance, quantiles (percentiles), and/or
the entire probability density function.

To address these open issues, we propose a Bayesian nonparametric regression
model (Karabatsos \&\ Walker, 2012\nocite{KarabatsosWalker12c}) for causal
inference in RDDs. It is an infinite-mixture model that allows the entire
probability density of the outcome variable to change flexibly as a function
of covariates. Our model can provide inferences of causal effects in terms
of how the treatment variable impacts the mean, variance, a quantile,
probability density function (p.d.f.), distribution function, and any other
chosen function of the outcome variable. Finally, the accurate estimation of
causal effects relies on an appropriate model for the data. Karabatsos and
Walker (2012\nocite{KarabatsosWalker12c}) showed that their Bayesian
nonparametric regression model tended to have better predictive performance
than other parametric and flexible nonparametric regression models of common
usage, over many real data sets.

Also, our model can be extended to handle causal inferences from a fuzzy
RDD\ (Trochim, 1984\nocite{Trochim84}). In contrast to a standard "sharp"\
RDD, a fuzzy RDD involves a study where not all subjects adhere to the
treatment assignment rule. This is because, for example, some subjects do
not comply with their respective treatment assignments, or because some
subjects receive treatments for which they are not eligible.

In Section \ref{Section ID RD}, we review the data assumptions that are
required to identify and estimate causal effects from a RDD. Unlike all
previous guides to performing causal inference from RDDs (e.g., Imbens \&\
Lemieux, 2008\nocite{ImbensLemieux2008}; Lee \&\ Lemieux, 2010\nocite%
{LeeLemieux10}; Bloom, 2012\nocite{Bloom12}; Wong et al. 2013\nocite%
{WongSteinerCook13}), we do not rely on the potential outcomes\ approach.
Instead we focus on the extended conditional independence approach, which
addresses the problem of causal inference entirely by the concepts of
standard probability theory (Dawid, 2000\nocite{Dawid00}, 2002\nocite%
{Dawid02}). We should also mention that we will not address SUTVA, as this
assumption only makes sense in the potential outcomes framework (Dawid, 2000%
\nocite{Dawid00}).

In Section 3, we describe our Bayesian nonparametric model that can estimate
causal effects from the various RDDs. In Section \ref{Section Apps}, we
illustrate our model through the analysis of two educational data sets,
involving (resp.) a sharp RDD and a fuzzy RDD. Section \ref{Section
Conclusions} concludes with a short discussion of the free user-friendly
software that can be used to implement our Bayesian nonparametric approach
to RDDs, and a discussion of possible extensions of our approach to
multivariate RDDs, involving cutoffs in more than one dimension.

Throughout, we denote by $\Pr (X\in A)$ the probability of an event $A$, for
a given random variable, $X$. Also, we assume that a continuous (resp.,
discrete) random variable $X$ admits a cumulative distribution function
(c.d.f.), denoted by a capital letter, such as $F(x)=\Pr (X\leq x),$ with
corresponding probability density (resp., mass) function\ or p.d.f., denoted
by a lower case letter, with $f(x)=\tint \mathrm{d}F(x)$. We use the
notation $X\sim F(x)$, $X\sim F$, or $X\sim f(x)$ to refer to $X$ as having
the distribution $F$. Accordingly, we denote by \textrm{n}$(\cdot \,|\,\mu
,\sigma ^{2})$ as the density of the normal \textrm{N}$(\cdot \,|\,\mu
,\sigma ^{2})$ c.d.f. with mean and variance $(\mu ,\sigma ^{2})$; with $%
\Phi \left( \cdot \right) =$ \textrm{N}$(\cdot \,|\,0,1)$ the c.d.f. of the
normal \textrm{n}$(\cdot \,|\,0,1)$ p.d.f.; $\mathrm{ga}(\cdot \,|\,a,b)$
and $\mathrm{ig}(\cdot \,|\,a,b)$ (resp.) denotes the p.d.f.s of the gamma $%
\mathrm{Ga}(\cdot \,|\,a,b)$ distribution and inverse gamma $\mathrm{IG}%
(\cdot \,|\,a,b)$ distribution (c.d.f.), with shape parameter $a$ and rate
parameter $b$; and $\mathrm{U}(\cdot \,|\,a,b)$ is the c.d.f. of a uniform
distribution.

\section{\textbf{Identifying Causal Effects in a RDD}\label{Section ID RD}}

A non-randomized study from a RDD involves three variables that are
observable from each individual from a sample of $n$ subjects, indexed by $%
i=1,\ldots ,n$. They are the \textit{outcome variable}, $Y$; a binary 
\textit{treatment variable} $T$, where $T=1$ refers to treatment \textit{%
receipt} and $T=0$ refers to non-treatment receipt; and a continuous-valued 
\textit{assignment variable} $R$. Each subject $i$\ is assigned the
treatment whenever $R_{i}\geq r_{0}$, and is assigned the non-treatment
whenever $R_{i}<r_{0}$, given a known fixed cutoff $r_{0}$. As will be
further described below, in a fuzzy RDD, being assigned treatment does not
imply receiving treatment. The treatment assignment variable is thus $%
\mathbf{1}_{R\geq r_{0}}$, with $\mathbf{1}_{(\cdot )}$ the indicator
function. A RDD study gives rise to a sample data set, $\mathcal{D}%
_{n}=\{(r_{i},t_{i},y_{i})\}_{i=1}^{n}$, including derived observations $%
\mathbf{1}_{r_{i}\geq r_{0}}$, and possibly observations of $p$ pretreatment
covariates, $\mathbf{x}_{i}=(x_{1i},\ldots ,x_{pi})^{\top }$. Also, we
introduce a non-random regime parameter, $\Psi _{T}\in \{\emptyset ,0,1\}$.
An "idle" or "observational" regime is indicated by $\Psi _{T}=\emptyset $,
when the joint distribution of $(R,T,Y)$ "arises naturally." An intervention
regime is indicated by setting $\Psi _{T}=0$ or $1$. Here, $\Psi _{T}=1$
indicates the treatment intervention, and $\Psi _{T}=0$ indicates the
nontreatment intervention. The observational regime is none other than the
RDD\ itself, whereas the intervention regimes are hypothetical.

A characterizing assumption of the RDD is that the conditional probability
of treatment receipt is discontinuous at $r_{0}$. That is:%
\begin{equation}
\text{\textbf{Assumption RD: }}\lim_{r\downarrow r_{0}}\mathrm{\Pr }%
(T=1\,|\,r)\neq \lim_{r\uparrow r_{0}}\mathrm{\Pr }(T=1\,|\,r).  \label{RD}
\end{equation}

There are two types of RDDs. In the classical, \textit{sharp RDD}
(Thistlewaite \&\ Campbell, 1960\nocite{ThistlewaiteCampbell60}; Cook, 2008%
\nocite{Cook08}), the probability function $\Pr (T=1\,|\,r)$ has a
discontinuous jump of size 1 at $R=r_{0}$, with point mass probability
function:%
\begin{equation}
f(t\,|\,r)=\Pr (T=t\,|\,R=r)=\mathbf{1}_{r\geq r_{0}}^{t}(1-\mathbf{1}%
_{r\geq r_{0}})^{1-t}.  \label{Den T SRD}
\end{equation}%
Then the treatment receipt is identical to treatment assignment, with $T=%
\mathbf{1}_{R\geq r_{0}}$.

In the \textit{fuzzy RDD} (Trochim, 1984\nocite{Trochim84}), $\Pr
(T=1\,|\,R=r)$ has a discontinuous jump that is smaller than 1 at $r_{0}$.
Then $f(t\,|\,r)$ is not a point mass density. The smaller jump results from
imperfect treatment adherence (e.g., treatment non-compliance), where some
of the subjects of the given study were either assigned $\mathbf{1}_{r\geq
r_{0}}=0$ but received $T=1$, or assigned $\mathbf{1}_{r\geq r_{0}}=1$ but
received $T=0$.

For either type of RDD, a typical measure of the causal effect is defined by
the difference of conditional means (expectations) of $Y$ at $R=r_{0}$: 
\begin{subequations}
\label{SRD E est}
\begin{eqnarray}
\tau &=&\lim_{r\downarrow r_{0}}\mathbb{E}(Y\,|\,R=r,\Psi
_{T}=1)-\lim_{r\uparrow r_{0}}\mathbb{E}(Y\,|\,R=r,\Psi _{T}=0)
\label{SRD E est1} \\
&=&\mathbb{E}(Y\,|\,R=r_{0},\Psi _{T}=1)-\mathbb{E}(Y\,|\,R=r_{0},\Psi
_{T}=0)  \label{SRD E est2}
\end{eqnarray}%
where $\mathbb{E}(Y\,|\,r,t)=\tint y\,\mathrm{d}F(y\,|\,r,t)$.\ Also if $%
r_{0}^{+}=\lim_{r\downarrow r_{0}}r$, $r_{0}^{-}=\lim_{r\uparrow r_{0}}r$,
and $r_{0}^{+}=r_{0}^{-}=r_{0}$, then $R$ is continuous. A motivation for
restricting to $R=r_{0}$ in the definition of the causal effect (\ref{SRD E
est2}), is that this is the only effect that can be directly estimated from
given data ($\mathcal{D}_{n}$) of a RDD.

In general, for any choice of function $H\{\cdot \}$ of $Y$, the causal
effect is given by: 
\end{subequations}
\begin{subequations}
\label{Causal Eff}
\begin{eqnarray}
\tau _{H} &=&\lim_{r\downarrow r_{0}}\mathbb{E}(H\{Y\}\,|\,r,\Psi
_{T}=1)-\lim_{r\uparrow r_{0}}\mathbb{E}(H\{Y\}\,|\,r,\Psi _{T}=0) \\
&=&\mathbb{E}(H\{Y\}\,|\,r_{0},\Psi _{T}=1)-\mathbb{E}(H\{Y\}\,|\,r_{0},\Psi
_{T}=0).
\end{eqnarray}%
Therefore, depending on the choice of function $H\{\cdot \}$, causal effects
are not only interpretable in terms of the mean of $Y$\ (when $H\{Y\}=Y$),
but also in terms of the variance ($H\{Y\}=\{Y-\mathbb{E}(Y\,|\,r,t)\}^{2}$%
), cumulative distribution function (c.d.f.) $F(y\,|\,r,t)$ ($H\{Y\}=\mathbf{%
1}_{Y\leq y}$), probability density function (p.d.f.) ($f(y\,|\,r,t)$),
survival function $1-F(y\,|\,r,t)$, and so on. Inverting the c.d.f. obtains $%
F^{-1}(u\,|\,r,t)$, for $u\in \lbrack 0,1]$. Then causal effects can also be
interpreted in terms of the $u^{\text{th}}$ quantile of $Y$.

If $R$\ is discrete, then obviously $r^{+}\neq r^{-}\neq r_{0}$. Then
equation (\ref{Causal Eff}) (including (\ref{SRD E est})) still provides a
measure of causal effect, which may require additional extrapolation in its
estimation.

Next we describe how a causal effect $\tau _{H}$ is identified from the
sharp RDD.

\subsection{Identification in the Sharp RDD}

The sharp RDD can be characterized in terms of the extended conditional
independence framework of causal inference, extending the ideas from Dawid
(2002\nocite{Dawid02}, Sections 6.2, 7). In this RDD, it is typically
assumed that the joint p.d.f. of $(R,T,Y)$, conditionally on $\Psi _{T}=\psi
_{T}$, is given by: 
\end{subequations}
\begin{equation}
f(r,t,y\,|\,\psi _{T})=f(r)f(t\,|\,r,\psi _{T})f(y\,|\,r,t).
\label{SRD full joint}
\end{equation}%
This p.d.f. (\ref{SRD full joint}) gives rise to the conditional
independence properties (a) $R\perp \!\!\!\perp \Psi _{T}$ and (b) $Y\perp
\!\!\!\perp \Psi _{T}\,|\,R,T$, where $\perp \!\!\!\perp $ denotes
conditional independence (Dawid, 1979\nocite{Dawid79}). Property (a)\ states
that the distribution of $R$ is the same in both observational and
interventional circumstances. Meanwhile, (b)\ is a causal property, because
it says that the distribution of $Y$ is unaffected by the choice of
interventional regime $\Psi _{T}$, conditionally on $(R,T)$ (Dawid, 2002%
\nocite{Dawid02}). Properties (a) and (b)\ together imply that $R$ is a
sufficient covariate (Dawid, 2002\nocite{Dawid02}, 2010\nocite{Dawid10}).

Motivated by the causal property (b), we now focus on the conditional p.d.f.:%
\begin{equation}
f(t,y\,|\,r,\psi _{T})=f(t\,|\,r,\psi _{T})f(y\,|\,r,t).
\label{SRD full joint cond}
\end{equation}%
If the intervention parameter takes on a null value $\Psi _{T}=\emptyset $,
then the joint distribution of the random variables $(T,Y)$ arises
naturally. Then (\ref{SRD full joint cond}) reduces to the joint p.d.f. $%
f(t,y\,|\,r)=f(t\,|\,r)f(y\,|\,r,t)$, where $f(t\,|\,r)$ is the point mass
density (\ref{Den T SRD}).

In contrast, an intervention that sets $\Psi _{T}=t_{0}\in \{0,1\}$,
modifies $f(t\,|\,r,\psi _{T})$ to $\mathbf{1}(t=t_{0})$, in the joint
p.d.f. (\ref{SRD full joint cond}). Also, recall that the causal effect is
estimable only conditionally on $R=r_{0}$. Then the conditional p.d.f. of $Y$
can be written as:%
\begin{equation}
f(y\,|\,r_{0},t_{0})=f(y\,||\,r_{0},t_{0}),  \label{IntervDens SRD}
\end{equation}%
where $||$ denotes "conditioning by intervention" (Lauritzen, 2000\nocite%
{Lauritzen00}). The equality in (\ref{IntervDens SRD}) holds by virtue of
the causal property mentioned above. Then for a general choice of function $%
H\{\cdot \}$, the causal effect is given by a comparison of $\mathbb{E}%
(H\{Y\}\,|\,r_{0},t_{0}),$ for $t_{0}=0,1$, including the p.d.f.s $%
f(y\,|\,r_{0},t_{0})$. For example, the causal effect in terms of the
difference, as in the general definition (\ref{Causal Eff}).

Now we turn to the issue of identifying the causal effect from data. Suppose
that there is reason to believe that in the absence of treatment, subjects
close to the threshold $r_{0}$ are similar. Then for the sharp RDD, the
causal effect $\tau _{H}$\ is identified by assumption RD (\ref{RD}) and:%
\begin{equation}
\text{\textbf{Continuity at }}r_{0}\text{ for }\Psi _{T}=0\text{:\textbf{\ } 
}f(y\,|\,\,r,\,\Psi _{T}=0)\text{ is continuous\thinspace in}\,\,r\,\,\text{%
at}\,\,r_{0}\text{ for all }y\text{.}  \label{assump}
\end{equation}%
Equation (\ref{assump}) is a density version of the assumption in Hahn et
al. (2001\nocite{HahnToddVanderKlaauw01}), who only look at mean shifts and
hence assume that 
\begin{equation}
\mathbb{E}(Y\,|\,\,r,\,\Psi _{T}=0)\text{ is continuous\thinspace in}%
\,\,r\,\,\text{at}\,\,r_{0}.  \label{meanass}
\end{equation}%
We believe it is important to model mean shifts not by having a mean shift
model but rather by modeling the density of the observations and then
picking out the mean from this. This is the correct approach. Our
contribution involves replacing identifying assumption (\ref{meanass}) by (%
\ref{assump}), and hence we need to model the density nonparametrically.

Assuming (\ref{assump}) rather than only (\ref{meanass}) allows the
treatment effect to exhibit itself in more ways than a mean shift. For
example, a variance shift would also be informative, even in the absence of
a mean shift. The model we employ is quite general and allows many aspects
of treatment effect to be explored by studying any differences between
density estimate either side of the cut-off at $r_{0}$. However, we can
obviously estimate the key mean shift, having modeled the density functions
either side of the cut-off point $r_{0}$, simply by estimating the means of
the two density functions.

Authors such as Lee (2008\nocite{Lee08}) and Lee and\ Lemieux (2010\nocite%
{LeeLemieux10}) further elaborated on the continuity assumption (\ref{assump}%
). They showed that if subjects have \textit{imprecise control} of $R$ at $%
r_{0}$, then this continuity condition holds, and that treatments are "as
good as randomly assigned" for the subgroup of subjects having values of the
assignment variable $R$ located in a small neighborhood around $r_{0}$.

\subsection{Identification in the Fuzzy RDD}

In the fuzzy RDD, the probability function $\Pr (T=1\,|\,r)$ has a
discontinuous jump that is smaller than 1 at $r_{0}$, meaning that $R$ does
not determine $T$. Then $\mathbf{1}_{R\geq r_{0}}$ and $T$ are distinct
variables since the event $\mathbf{1}_{R\geq r_{0}}\neq T$ is possible; and $%
T$ and $Y$ may both depend on unobserved confounding variables, collectively
labelled as $U$. Given these considerations, we may extend the joint p.d.f. (%
\ref{SRD full joint}) to:%
\begin{equation}
f(r,t,y,u\,|\,\psi _{T})=f(r\,|\,u)f(t\,|\,r,u,\psi _{T})f(y\,|\,r,t,u)f(u).
\label{FRD full joint}
\end{equation}%
This p.d.f. admits the conditional independence properties (a)\ $R\perp
\!\!\!\perp \Psi _{T}\,|\,U$ and (b) $Y\perp \!\!\!\perp \Psi _{T}\,|\,R,T,U$
(and condition (a)\ can be strengthened to $(R,U)\perp \!\!\!\perp \Psi
_{T}\,|\,U$). Here, (b)\ is a causal property, and (a) and (b)\ together
imply the assumption that $R$ is a sufficient covariate, conditionally on $U$%
. But for the purposes of making causal inferences from real data, we cannot
condition on $(R,T,U)$ because $U$ is unobserved.

However, for the fuzzy RDD, the assumptions RD\ (\ref{RD}) and continuity (%
\ref{assump}), along with certain additional assumptions also imply that the
causal effect (\ref{Causal Eff}) is identified by the ratio:%
\begin{equation}
\tau _{H}=\dfrac{\lim_{r\downarrow r_{0}}\mathbb{E}(H\{Y\}\,|\,r)-\lim_{r%
\uparrow r_{0}}\mathbb{E}(H\{Y\}\,|\,r)}{\lim_{r\downarrow r_{0}}\mathbb{E}%
(T\,|\,r)-\lim_{r\uparrow r_{0}}\mathbb{E}(T\,|\,r)},  \label{IVest}
\end{equation}%
for a general function $H\{\cdot \}$ of $Y$. The numerator of (\ref{IVest})
is the Intention to Treat (ITT)\ effect. The denominator is a measure of
treatment adherence, which decreases as noncompliance increases. In the
sharp RDD, where $\mathbf{1}_{R\geq r_{0}}=T$, the denominator is 1 (i.e.,
perfect adherence), and then the ITT\ effect coincides with the causal
effect $\tau _{H}$. See Hahn et al. (2001\nocite{HahnToddVanderKlaauw01})
for more details.

\section{\textbf{Estimating Causal Effects In a RDD}}

Here, we propose our Bayesian nonparametric model for causal inference in
RDDs.

\subsection{Bayesian Nonparametric Model}

For the sharp RDD, our Bayesian nonparametric model (Karabatsos \&\ Walker,
2012\nocite{KarabatsosWalker12c}) is defined by: 
\begin{subequations}
\label{BNP model}
\begin{eqnarray}
f(y_{i}\,|\,r_{i},t_{i};\,\mathbf{\zeta }) &=&\dsum\limits_{j=-\infty
}^{\infty }\mathrm{n}(y_{i}\,|\,\mu _{j},\sigma _{j}^{2})\omega _{j}\{\eta
(r_{i},t_{i}),\sigma (r_{i},t_{i})\},\text{ \ }i=1,\ldots ,n,
\label{BNPlike} \\
\text{ \ \ \ \ \ \ \ \ \ \ \ \ \ \ }\omega _{j}(\eta ,\sigma ) &=&\Phi
\QDOVERD( ) {j-\eta }{\sigma }-\Phi \QDOVERD( ) {j-1-\eta }{\sigma }
\label{BNP weights} \\
\eta (r,t) &=&\beta _{0}+\beta _{1}r+\beta _{2}t  \label{etaterm} \\
\sigma ^{2}(r,t) &=&\exp (\lambda _{0}+\lambda _{1}r+\lambda _{2}t)
\label{lambdaterm} \\
\mu _{j},\sigma _{j}^{2}\,|\,\mu _{\mu },\sigma _{\mu }^{2},b_{\sigma }
&\sim &\mathrm{N}(\mu _{j}\,|\,\mu _{\mu },\sigma _{\mu }^{2})\mathrm{IG}%
(\sigma _{j}^{2}\,|\,1,b_{\sigma }),\text{ }j=0,\pm 1,\pm 2,\ldots \\
\mu _{\mu },\sigma _{\mu }^{2} &\sim &\mathrm{N}(\mu _{\mu }\,|\,\mu
_{0},\sigma _{0}^{2})\mathrm{U}(\sigma _{\mu }\,|\,0,b_{\sigma \mu }) \\
b_{\sigma },\mathbf{\beta },\boldsymbol{\lambda } &\sim &\mathrm{Ga}%
(b_{\sigma }\,|\,a_{0},b_{0})\mathrm{N}(\mathbf{\beta },\mathbf{\lambda }%
\,|\,\mathbf{0,}v\mathbf{I})
\end{eqnarray}%
where the mixture weights $\omega _{j}\{\eta (r,t),\sigma (r,t)\}$ sum to 1
at each value of $(r,t)$. Also, the terms (\ref{BNPlike}) and (\ref{BNP
weights}) may be deconstructed via the generation of a latent indicator
variable $Z\sim $ \textrm{N}$(\eta ,\sigma ^{2}),$ and then taking $Y\sim 
\mathrm{N}(\mu _{j},\sigma _{j}^{2})$ if $j-1<Z\leq j$.

The model (\ref{BNP model}) allows the entire probability density of the
outcome variable $Y$ to change flexibly as a function of covariates. The
parameter $\sigma (r,t)$ measures the multimodality of $f(y\,|\,r,t)$
(Karabatsos \&\ Walker, 2012\nocite{KarabatsosWalker12c}). Specifically, as $%
\sigma (r)\rightarrow \infty $, the density $f(y\,|\,r,t)$ becomes more
multimodal, with weights $\omega _{j}\{\eta (r,t),\sigma (r,t)\}$ converging
to a discrete uniform distribution; and as $\sigma (r,t)\rightarrow 0$, the
density $f(y\,|\,r,t)$ becomes more unimodal, and "local," with $%
f(y\,|\,r,t)\approx \mathrm{n}(y_{i}\,|\,\mu _{j},\sigma _{j}^{2})$ and 
\end{subequations}
\begin{equation*}
\omega _{j}\{\eta (r,t),\sigma (r,t)\}\approx 1\text{ \ if }j-1<\eta \leq j.
\end{equation*}%
Furthermore, the model has a discontinuity at $r_{0}$ due to the presence of
the term $T$ in both (\ref{etaterm}) and (\ref{lambdaterm}). The effect,
controlled by the coefficients $(\mathbf{\beta },\boldsymbol{\lambda })$, is
to reallocate the weights either side of $r_{0}$, resulting in different
densities either side of this value. Obviously, there is a discontinuity if
and only if either of the coefficients $(\lambda _{2},\beta _{2})$ is
non-zero. The normal prior $\mathrm{N}(\mathbf{\beta },\mathbf{\lambda }\,|\,%
\mathbf{0,}v\mathbf{I})$ consists of a prior variance parameter $v$, which
controls for both the prior support for the range of the mixture density
component indices $j=0,\pm 1,\pm 2,\ldots $ (via the parameter $\mathbf{%
\beta }$), and for the range of the level of multimodality in $f(y\,|\,r,t)$%
. As $v\rightarrow \infty $, a wider range of component densities and
multimodality is supported; and as $v\rightarrow 0$, $f(y\,|\,r,t)$ becomes
a normal density.

When prior information is limited about the model parameters, we may attempt
to specify non-informative priors, for example, by choosing $\mu
_{0}=0,\sigma _{0}^{2}\rightarrow \infty $, $a_{0}\rightarrow 0$, $%
b_{0}\rightarrow 0$, and $v=10^{5}$, and by choosing $b_{\sigma \mu }$
according to prior knowledge about range of the $Y$ variance. For instance,
if $Y$ is known to have a variance of $1$, then $b_{\sigma \mu }=5$ provides
a vague prior choice. For such choices of prior parameters, the Bayesian
model (\ref{BNP model}), over 22 real data sets, demonstrated very good
predictive accuracy, and better predictive accuracy compared to many other
regression models, and compared to the Bayesian model under different
choices of prior (Karabatsos \&\ Walker, 2012\nocite{KarabatsosWalker12c}).

The model (\ref{BNP model}) has infinite-dimensional parameter, $\mathbf{%
\zeta }=((\mu _{j},\sigma _{j}^{2})_{j=-\infty }^{\infty },$ $\mu _{\mu },$ $%
\sigma _{\mu }^{2},$ $b_{\sigma },$ $\mathbf{\beta },$ $\boldsymbol{\lambda }%
)$, with prior density $\pi (\mathbf{\zeta })$. A set of data $\mathcal{D}%
_{n}=\{(y_{i},r_{i},t_{i})\}_{i=1}^{n}$ updates the prior $\pi (\mathbf{%
\zeta })$ to a posterior density, given by 
\begin{equation*}
\pi (\mathbf{\zeta }\,|\,\mathcal{D}_{n})=\dfrac{\tprod%
\nolimits_{i=1}^{n}f(y_{i}\,|\,r_{i},t_{i};\mathbf{\zeta })\pi (\mathbf{%
\zeta })}{\dint \tprod\nolimits_{i=1}^{n}f(y_{i}\,|\,r_{i},t_{i};\mathbf{%
\zeta })\mathrm{d}\Pi (\mathbf{\zeta })},
\end{equation*}%
with $\Pi (\mathbf{\zeta })$ (and $\Pi (\mathbf{\zeta }\,|\,\mathcal{D}_{n})$%
, resp.) the c.d.f. of $\pi (\mathbf{\zeta })$ (of $\pi (\mathbf{\zeta }\,|\,%
\mathcal{D}_{n})$). Also, let $F(y\,|\,r,t;\mathbf{\zeta })$ be the c.d.f.
of $f(y\,|\,r,t;\mathbf{\zeta })$. Then the posterior predictive density, $%
f_{n}(y\,|\,r,t)$, and the conditional posterior predictive expectation ($%
\mathbb{E}_{n}$) and variance ($\mathbb{V}_{n}$) of the outcome $H\{Y\}$ are
given (resp.) by:%
\begin{eqnarray*}
f_{n}(y\,|\,r,t) &=&\dint f(y\,|\,r,t;\,\mathbf{\zeta })\mathrm{d}\Pi (%
\mathbf{\zeta }\,|\,\mathcal{D}_{n}), \\
\mathbb{E}_{n}(H\{Y\}\,|\,r,t) &=&\dint \{\tint H\{y\}\mathrm{d}F(y\,|\,r,t;%
\mathbf{\zeta })\}\mathrm{d}\Pi (\mathbf{\zeta }\,|\,\mathcal{D}_{n}), \\
\mathbb{V}_{n}(H\{Y\}\,|\,r,t) &=&\dint \left[ \tint \{H\{y\}-\mathbb{E}%
_{n}(H\{Y\}\,|\,r,t)\}^{2}\mathrm{d}F(y\,|\,r,t;\,\mathbf{\zeta })\right] 
\mathrm{d}\Pi (\mathbf{\zeta }\,|\,\mathcal{D}_{n}).
\end{eqnarray*}%
Depending on the choice of function $H\{\cdot \}$, the posterior mean $%
\mathbb{E}_{n}$\ and variance $\mathbb{V}_{n}$ of the conditional
expectation $\mathbb{E(}Y\,|\,r,t)$, variance $\mathbb{V(}Y\,|\,r,t)$,
c.d.f. $F\mathbb{(}y\,|\,r,t)$ at a point $y$, are given (resp.)\ by $%
\mathbb{E}_{n}\{\mathbb{E(}Y\,|\,r,t)\}$ and $\mathbb{V}_{n}\{\mathbb{E(}%
Y\,|\,r,t)\}$; $\mathbb{E}_{n}\{\mathbb{V(}Y\,|\,r,t)\}$ and $\mathbb{V}%
_{n}\{\mathbb{V(}Y\,|\,r,t)\}$; and $\mathbb{E}_{n}\{F\mathbb{(}%
y\,|\,r,t)\}=F_{n}\mathbb{(}y\,|\,r,t)$ and $\mathbb{V}_{n}\{F\mathbb{(}%
y\,|\,r,t)\}$. For assessing the fit of the Bayesian model to data, a
standardized residual for each observation $y_{i}$ may be computed by%
\begin{equation*}
\overline{z}_{i}=\{y_{i}-\mathbb{E}_{n}(Y\,|\,r_{i},t_{i})\}/\{\mathbb{V}%
_{n}(Y\,|\,r_{i},t_{i})\}^{1/2}.
\end{equation*}%
If $|\overline{z}_{i}|>2,$ then $y_{i}$ can be judged as an outlier.

\subsection{Estimating Causal Effects with the Bayesian Model}

For our Bayesian model, the posterior estimates of the causal effect of $T$
on $H\{Y\}$, conditionally on $R=r_{0}$, are given as follows, under the
assumptions RD (\ref{RD}) and continuity at $r_{0}$ (\ref{assump}).

For the sharp RDD, the estimate $\widehat{\tau }_{H}$\ of the causal effect
is given by%
\begin{equation}
\widehat{\tau }_{H}^{(S)}=\mathbb{E}_{n}(\tau _{H}^{(S)})=\mathbb{E}%
_{n}(H\{Y\}\,|\,r_{0},T=1)-\mathbb{E}_{n}(H\{Y\}\,|\,r_{0},T=0),
\label{BayesSharpEst}
\end{equation}%
with posterior variance $\mathbb{V}_{n}(\tau _{H}^{(S)})=\mathbb{V}%
_{n}(H\{Y\}\,|\,r_{0},1)+\mathbb{V}_{n}(H\{Y\}\,|\,r_{0},0)$. Then $\widehat{%
\tau }_{H}^{(S)}\pm 2[\mathbb{V}_{n}(\tau _{H}^{(S)})]^{1/2}$ provides an
approximate 95\% posterior confidence band around $\widehat{\tau }_{H}^{(S)}$%
.

When inferring the causal effect in terms of the $u^{th}$ quantile, via $%
\widehat{\tau }_{H}^{(S)}=F_{n}^{-1}\mathbb{(}u\,|\,r,1)-F_{n}^{-1}\mathbb{(}%
u\,|\,r,0),$ we may judge whether $\widehat{\tau }_{H}$ is significantly
different from zero by using a P-P\ plot (Wilk \&\ Gnanadesikan, 1968\nocite%
{WilkGnanadesikan68})\ to check for non-overlap of the 95\%\ posterior
credible intervals ($F_{n(.025)}\mathbb{(}y\,|\,r,t),F_{n(.975)}\mathbb{(}%
y\,|\,r,t))$ at $u,$ for $T=0,1$, and over a wide range of points $y\in 
\mathcal{Y}$. Here, $F_{n(.025)}\mathbb{(}y\,|\,r,t)$ ($F_{n(.975)}\mathbb{(}%
y\,|\,r,t),$ resp.) denotes the posterior 2.5th percentile (97.5th
percentile, resp.) of $F\mathbb{(}y\,|\,r,t)$.

For the fuzzy RDD, the causal effect estimate $\widehat{\tau }_{H}$, in
terms of the ratio estimator (\ref{IVest}), may be obtained by two
independent regressions. The first involves estimating the numerator using
our regression model (\ref{BNP model}), after replacing the covariate $T$
with $\mathbf{1}_{R\geq r_{0}}$. The second involves a regression of $T$ on $%
(R,\mathbf{1}_{R\geq r_{0}})$ to estimate the denominator, via the posterior
predictive expectations:%
\begin{equation*}
\mathbb{E}_{n}(T\,|\,r,\mathbf{1}_{r\geq r_{0}}=a)=\tint \Pr (T=1\,|\,r,a;\,%
\mathbf{\zeta }_{T})\mathrm{d}\Pi (\mathbf{\zeta }_{T}\,|\,\mathcal{D}_{n}),%
\text{ }a=0,1.
\end{equation*}%
Our model (\ref{BNP model}) can be extended to binary regression, by
modeling the response density by:%
\begin{multline}
\Pr (T_{i}=1\,|\,r_{i},\mathbf{1}_{r_{i}\geq r_{0}};\,\mathbf{\zeta }_{T})
\label{BinaryOut} \\
=\dint\limits_{0}^{\infty }\left[ \tsum\limits_{j=-\infty }^{\infty }\mathrm{%
n}(t_{i}^{\ast }\,|\,\mu _{j},\sigma _{j}^{2})\omega _{j}\{\eta (r_{i},%
\mathbf{1}_{r_{i}\geq r_{0}}),\sigma (r_{i},\mathbf{1}_{r_{i}\geq r_{0}})\}%
\right] \mathrm{d}t_{i}^{\ast },\text{\ \ }i=1,\ldots ,n\text{.}
\end{multline}%
analogous to (\ref{BNPlike}). This provides flexible modeling of the inverse
link function by a covariate dependent, infinite mixture of normal c.d.f.s
(Karabatsos \&\ Walker, 2012\nocite{KarabatsosWalker12c}).

As before, denote $\pi (\mathbf{\zeta }\,|\,\mathcal{D}_{n})$ as the
posterior density for the model (\ref{BNP model}) for the $Y$ outcome; and
denote $\pi (\mathbf{\zeta }_{T}\,|\,\mathcal{D}_{n})$ as the posterior
density for the version of the model for the $T$ outcome, using (\ref%
{BinaryOut}). Then both posterior densities admit the conditional
independence property $\mathbf{\zeta \perp \!\!\!\perp \zeta }_{T}\,|\,%
\mathcal{D}_{n}$, so then we can write $\pi (\mathbf{\zeta },\mathbf{\zeta }%
_{T}\,|\,\mathcal{D}_{n})=\pi (\mathbf{\zeta }\,|\,\mathcal{D}_{n})\pi (%
\mathbf{\zeta }_{T}\,|\,\mathcal{D}_{n})$. This means that the posterior
densities of both models, $\pi (\mathbf{\zeta }\,|\,\mathcal{D}_{n})$ and $%
\pi (\mathbf{\zeta }_{T}\,|\,\mathcal{D}_{n})$, can be estimated either
separately or jointly.

For the fuzzy RDD, an estimate of the causal effect, in terms of the ratio (%
\ref{IVest}), is given by the posterior average of the ratio:%
\begin{equation*}
\mathbb{E}_{n}(\tau _{H}^{(F)})=\dint \left\{ \dfrac{\mathbb{E}%
(h\{Y\}\,|\,r_{0},1;\,\mathbf{\zeta })-\mathbb{E}(h\{Y\}\,|\,r_{0},0;\,%
\mathbf{\zeta })}{\mathbb{E}(T\,|\,r_{0},1;\,\mathbf{\zeta }_{T})-\mathbb{E}%
(T\,|\,r_{0},0;\,\mathbf{\zeta }_{T})}\right\} \pi (\mathbf{\zeta },\mathbf{%
\zeta }_{T}\,|\,\mathcal{D}_{n})\mathrm{d}(\mathbf{\zeta },\mathbf{\zeta }%
_{T}),
\end{equation*}%
for a given choice of function $h\{\cdot \}$, where the differences in the
ratio above are based on values of $\mathbf{1}_{r\geq r_{0}}=0,1$. A
computationally fast (but somewhat ad-hoc) first-order Taylor approximation
to $\mathbb{E}_{n}(\tau _{H}^{(F)})$ is given by:%
\begin{equation}
\widehat{\tau }_{H}^{(F)}=\widehat{\tau }_{H}^{(S)}/\{\mathbb{E}%
_{n}(T\,|\,r_{0},\mathbf{1}_{r\geq r_{0}}=1)-\mathbb{E}_{n}(T\,|\,r_{0},%
\mathbf{1}_{r\geq r_{0}}=0)\}=\widehat{\tau }_{H}^{(S)}/\mathbb{E}%
_{n}(D_{T}).  \label{BayesIVest}
\end{equation}%
For example, given the choice of function $H\{Y\}=\mathbf{1}_{Y\leq y}$, we
have the causal effect defined by a comparison of c.d.f.s at a point $y$,
weighted by $\mathbb{E}_{n}(D_{T})$, with 
\begin{equation*}
\widehat{\tau }_{\mathbf{1(}Y\leq y)}^{(F)}=\{F_{n}\mathbb{(}y\,|\,r,1)-F_{n}%
\mathbb{(}y|\,r,0)\}/\mathbb{E}_{n}(D_{T}).
\end{equation*}%
The second-order approximation is given by%
\begin{equation*}
\widehat{\tau }_{H}^{(F[2])}=\{\widehat{\tau }_{H}^{(S)}/\mathbb{E}%
_{n}(D_{T})\}+[\{\widehat{\tau }_{H}^{(S)}\mathbb{V}(D_{T})\}/\{\mathbb{E}%
_{n}(D_{T})\}^{3}\}],
\end{equation*}%
with $\mathbb{V}(D_{T})=\mathbb{V}_{n}(T\,|\,r_{0},1)+\mathbb{V}%
_{n}(T\,|\,r_{0},0)$. The posterior variance $\mathbb{V}_{n}(\tau
_{H}^{(F)}) $\ has first-order approximation:%
\begin{equation}
\mathbb{V}_{n}(\tau _{H}^{(F)})\approx \{\widehat{\tau }_{H}^{(S)}/\mathbb{E}%
_{n}(D_{T})\}^{2}[\mathbb{V}_{n}(\tau _{H}^{(S)})(\widehat{\tau }%
_{H}^{(S)})^{-2}+\mathbb{V}(D_{T})\{\mathbb{E}_{n}(D_{T})\}^{-2}].
\label{VarFuzzy}
\end{equation}%
These approximations are derived from standard results involving the
distribution of the ratio of two random variables (e.g., Stuart \&\ Ord, 1998%
\nocite{StuartOrd98}, p. 351). Then $\widehat{\tau }_{H}^{(F)}\pm 2\{\mathbb{%
V}_{n}(\tau _{H}^{(F)})\}^{1/2}$ gives a 95\% posterior interval around $%
\widehat{\tau }_{H}^{(F)}$. Also, when inferring the causal effect $\widehat{%
\tau }_{H}^{(F)}$\ in terms of treatment and non-treatment differences at
the $u^{th}$ quantile, we may judge whether $\widehat{\tau }_{H}$ is
significantly different from zero by using a P-P\ plot of the 95\% posterior
intervals $F_{n}\mathbb{(}y\,|\,r,a)\pm 2\{\mathbb{V}_{n}(\tau _{\mathbf{1(}%
Y\leq y)}^{(F)})\}$, over points $y\in \mathbb{R}$, and then checking for
nonoverlap for these intervals at point $u$.

Alternatively, it may be of interest to investigate the sensitivity of the
causal effect estimate $\widehat{\tau }_{H}^{(F)}$ to variations of
treatment adherence (e.g., compliance). This can be achieved by estimating
the ratio $\widehat{\tau }_{H}^{(F)}$for each of a set of fixed non-zero
values (e.g., $1,.9,.8,...,-1$) for the denominator, with each estimate
having posterior variance%
\begin{equation*}
\mathbb{V}_{n}(\tau _{H}^{(F)})\approx (\widehat{\tau }_{H}^{(S)}/\mathbb{E}%
_{n}(D_{T}))^{2}\{(\mathbb{V}_{n}(\tau _{H}^{(S)}))(\widehat{\tau }%
_{H}^{(S)})^{-2}\}.
\end{equation*}

Using Markov Chain Monte Carlo (MCMC), Gibbs sampling methods, along with a
slice sampling step for $\sigma _{\mu }$, can be used to estimate all of the
aforementioned posterior quantities (Karabatsos \&\ Walker, 2012\nocite%
{KarabatsosWalker12c}). We use Rao-Blackwell (RB) methods to estimate all
the posterior linear functionals, such as $\mathbb{E}_{n}(H\{Y\}\,|\,r,a)$, $%
\mathbb{V}_{n}(H\{Y\}\,|\,r,a)$, $r_{i}$, $\mathbb{E}_{n}(T\,|\,r,a)$ (for $%
a=0,1$), $\mathbb{V}_{n}(T\,|\,r_{0},a)$, $\mathbb{E}_{n}(\tau _{H}^{(S)})$, 
$\mathbb{V}_{n}(\tau _{H}^{(S)})$, $\mathbb{E}_{n}(\tau _{H}^{(F)})$, and $%
\mathbb{V}_{n}(\tau _{H}^{(F)})$ (Gelfand \&\ Mukhopadhyay, 1995\nocite%
{GelfandMukhopadhyay95}).

\section{\textbf{Illustrative Applications}\label{Section Apps}}

The Bayesian nonparametric model was illustrated through the analysis of two
data sets, using menu-driven software that was developed by the first author
(Karabatsos, 2014,a,b\nocite{Karabatsos14a}\nocite{Karabatsos14b}). The
first data set was collected from four Chicago University schools of
education, which established a new curriculum that aims to train teachers to
help improve Chicago public schools. This data set involved a sharp RDD. The
second data set, obtained from Angrist and Lavy (2008\nocite{AngristLavy08}%
), involves a fuzzy RDD, from a study of the effect of class size on student
achievement (Angrist \&\ Lavy, 1999\nocite{AngristLavy99}). For each data
set, it seems reasonable to make the assumptions of RD\ (\ref{RD}) and
continuity (\ref{assump}) at $r_{0}$ (i.e., imprecise control at $r_{0}$),
in order to identify the causal effects of treatment on the outcome,
conditionally on $r_{0}$ (see Section 2).

For both data sets, the Bayesian nonparametric model assumed the same vague
priors that were mentioned in Section 3.2. According to standardized
residuals, the model under these priors provided good predictive accuracy
for each data set.

All posterior estimates of this model, reported in the next two subsections,
are based on 40K MCMC\ samples. These samples were obtained from every 5th
iterate of a run of 200K\ MCMC\ sampling iterations, after discarding the
first 2K\ burn-in samples. This provided accurate posterior estimates
according to standard convergence assessments (Geyer, 2011\nocite{Geyer11}).
Specifically, univariate trace plots displayed good mixing of model
parameters and posterior predictive samples, and all posterior predictive
estimates obtained 95\%\ MC\ confidence intervals with half-width sizes near
.01.

\subsection{Learning Math Teaching:\ Time Series Data}

For the first data set, the aim is to estimate the effect of the new teacher
education curriculum on math teaching ability, among $n=347$ undergraduate
teacher education students attending one of four Chicago universities. This
data set involves a sharp RDD, an interrupted time-series design (Cook \&\
Campbell, 1979\nocite{CookCampbell79}) using an assignment variable of time,
ranging from fall semester 2007 through spring semester 2013. The new
curriculum (treatment)\ was instituted in Fall 2010 (the cutoff, $r_{0}$),
and the old teacher curriculum (non-treatment) was active before then. The
outcome variable ($Y$) is the number-correct score on the 25-item Learning
Math for Teaching (LMT) test (LMT, 2012\nocite{LMT12}). Each of the students
completed the LMT\ test ($89.9\%$ female; $135$ and $212$ students under the
old and new curriculum), after finishing a course on teaching algebra. Among
them, the average LMT\ score was $12.9$ (s.d. $=3.44$), with Cronbach's
alpha reliability $.63$. The LMT\ scores were transformed to z-scores,
having sample mean 0 and variance 1.

Using our Bayesian model, we analyzed the data to estimate the effect of the
new curriculum, versus the old curriculum, on student ability to teach math
(LMT\ score), at the Fall semester 2010 cutoff. The model included the LMT\
test z-score as the\ dependent variable ($Y$), and included covariates of
the assignment variable ($R$), given by $\mathrm{TimeF10}=\mathrm{Year}%
-2010.6$, and of the treatment assignment variable \textrm{CTPP} $=\mathbf{1}%
_{\mathrm{Year}\geq 2010.6}$. The cutoff $2010.6$ is the time midpoint
between Spring 2010 ($2010.3$)\ and Fall 2010 ($2010.9$).

For the model, R-squared was $.99$, and nearly all the standardized
residuals ranged between $-1$ and $1$ over the $347$ observations, with one
residual slightly exceeding 2. Figure 1 presents the model's posterior
predictive density estimate of the LMT\ outcome, for the new curriculum
(treatment) and for the old curriculum (non-treatment), at Fall 2010. The
new curriculum, compared to the old, increased the LMT\ scores, by shifting
the density of LMT scores to the right. This shift corresponds to an
increase in the mean (from $-.17$ to $-.13$), the 10th percentile ($-2.01$
to $-1.97$), and the 25th percentile ($-1.43$ to $-1.31$), but these
increases were not statistically significant from zero according to 95\%\
credible intervals of the predictive mean and of the posterior c.d.f.
estimates. Also, each density presents two modes (clusters) of students,
indicating the presence of a latent binary covariate.

\begin{center}
-----------------------

Insert Figure 1

-----------------------
\end{center}

\subsection{Maimonides' Data:\ Fuzzy RDD}

The twelfth-century rabbinic scholar Maimonides proposed a rule that
specifies a maximum class size of 40, under the belief that smaller class
sizes promotes higher student achievement (see Hyamson, 1937\nocite%
{Hyamson37}, p. 58b). Specifically, for a given class $c$ in school $s$, the
rule assigns average class size (\textrm{Psize}$_{sc}$) as a function of
beginning-of-the-year school enrollment ($e_{s}$), according to the
prediction equation \textrm{Psize}$_{sc}=e_{s}/\mathrm{floor}%
[((e_{s}-1)/40)+1]$. The rule (equation) assigns students of a school into a
single classroom when the school's enrollment is less than $41$, assigns
students into two classrooms of average size $20.5$ when school enrollment
reaches $41$; assigns students into three classrooms of average size $27$
when enrollment reaches $81$; and so on. The cutoff number $20.5$
distinguishes between small and large classes.

Here, we study the effect of class size on average class verbal achievement,
through the analysis of data on 4th grade students who each attended one of
2,056 classes in Israeli public schools during 1991. These schools used
Maimonides' rule to allocate students into classrooms. Demographic
statistics are reported in Angrist and\ Lavy (1999\nocite{AngristLavy99})
(three other classes were not analyzed because they had missing achievement
data). For the Bayesian model, the dependent variable ($Y$) is average class
verbal score (\textrm{avgverb}), which we transformed to z-scores with
sample mean 0 and variance 1. The covariates include the assignment variable
($R$), defined by the rule-predicted class size centered at the cutoff $20.5$
(i.e., \textrm{Psize205 }$=$ \textrm{Psize }$-$ $20.5$), and include the
indicator of large (vs. small) class assignment, \textrm{Plarge }$=$ $%
\mathbf{1}_{\mathrm{Psize}\geq 20.5}$. Now, while Maimonides' rule may
assign a given class to be a large (small, resp.), the class could become
small (large, resp.). For example, one school in the data set had an
enrollment of 41, leading to some students receiving a large class of 21,
and other students receiving a small class of 20. Therefore, the data arise
from a fuzzy RDD, and for the data analysis, we also consider a variable
defined by the indicator of large class receipt, \textrm{large }$=$\textrm{\ 
}$\mathbf{1}_{\mathrm{classize}\geq 20.5}$. We also fit the Bayesian model,
with the treatment ($T$) variable, \textrm{large}, as the dependent
variable, and with covariates \textrm{Psize205} and \textrm{Plarge.}

For the \textrm{avgverb} ($Y$) dependent variable, the Bayesian model
obtained an R-squared of $.88$, with standardized fit residuals ranging from 
$-1.1$ to $1.3$ over the $2,056$ observations. Thus the model had no
outliers. For the treatment ($T$) dependent variable, \textrm{large}, the
Bayesian model had no outliers, and estimated $.93$ as the denominator of
the causal effect estimator (\ref{IVest}). Figure 2 presents the model's
posterior predictive density estimates of the \textrm{avgverb }outcomes, for
the treatment versus the non-treatment, each divided by $.93$. It was found
that large class size (versus small)\ causally increased the verbal score,
in terms of the 5th percentile ($-2.45$ to $-2.19$), 10th percentile ($-1.81$
to $-1.66$), 25th percentile ($-.94$ to $-.88$), and causally decreased the
score in terms of the 75th percentile ($.90$ to $.79$), 90th percentile ($%
1.48$ to $1.37$), and 95th percentile ($\allowbreak 1.76$ to $1.66$). Each
of these estimates is based on taking the predictive quantile estimates of
the Bayesian model for \textrm{avgverb} ($Y$), and dividing them by $.93$.
Also, each density presents two modes (clusters) of students, indicating the
presence of a latent binary covariate.

\begin{center}
-----------------------

Insert Figure 2

-----------------------
\end{center}

\section{Conclusions\label{Section Conclusions}}

We proposed and illustrated a flexible Bayesian nonparametric regression
model for causal inference in RDDs. Such designs identify causal effects
under relatively mild conditions. While the existing linear models for RDDs
only focus on mean causal effects, the Bayesian model provides inferences of
causal effects in terms of the mean, variance, distribution function,
quantile, probability density, or any other functional of the outcome
variable.

In future work, the Bayesian nonparametric regression modeling approach will
be extended to handle RDDs\ involving a multivariate assignment variable, $%
\mathbf{R\in 
\mathbb{R}
}^{K}$, which assigns to the treatment condition (versus nontreatment) if
and only if $\mathbf{R}\in S$, for some set $S$. In this case, the measure
of the causal effect ($\tau _{H}$) no longer depends on a single cutpoint $%
r_{0}$, but instead depends on multiple cutpoints, defined by the boundary
points of $S$. In principle, it is straightforward to extend the model to
handle a multivariate RDD, because then the model would simply include $%
\mathbf{R}$, along with with a 0-1 indicator of the event $\mathbf{R}\in S$,
as covariates. A future study will carefully study how causal effects can be
summarized over the multiple boundary points of $S$, via the model's
posterior predictive distribution.

For the Bayesian nonparametric model discussed in this chapter, a
user-friendly and menu-driven software is freely available, entitled:\
"Bayesian Regression: Nonparametric and Parametric Models" (Karabatsos,
2014a,b\nocite{Karabatsos14a}\nocite{Karabatsos14b}).\ This free software
package can be downloaded and installed from:%
\begin{equation*}
\text{http://tigger.uic.edu/\symbol{126}georgek/HomePage/BayesSoftware.html.}
\end{equation*}%
The Bayesian nonparametric model can be easily specified for data analysis,
by clicking the menu options "Specify New Model" and "Infinite probits
regression model." Afterwards, the item response (dependent) variable,
covariates, and prior parameters can be easily selected (clicked)\ by the
user. Then, to run for data analysis, the user clicks the "Run Posterior
Analysis" button to run the MCMC sampling algorithm for a chosen number of
sampling iterations. Immediately after the completion of the MCMC\ run, the
software automatically opens a text output file containing the results of
the data analysis, including summaries of the posterior distribution of the
model, obtained from the MCMC samples. The software also allows the user to
conveniently check for MCMC\ convergence, through a menu options that can be
clicked to construct trace plots, and through another menu option that can
be clicked to run a batch means analyses to construct 95\%\ Monte Carlo
confidence intervals of the posterior estimates of the model parameters.
Other menu options of the software allows the user to construct plots and
(additional)\ text output of the (MCMC\ estimated marginal)\ posterior
distributions of the model parameters (e.g., box plots), and allows the user
to output text and residual plots that report the fit of the model in
greater detail. Additional menu options allow the user to construct
posterior predictions of the model, as a function of the covariates, in
terms of the mean, variance, quantile, probability density, distribution
function, or other chosen functions of the outcome variable.

Currently, the software provide the user a choice of 59 statistical models,
including a large number of Bayesian nonparametric\ regression models. The
software allows the user to specify Dirichlet process (DP)\ mixture models,
and more generally, mixture regression models based on the stick-breaking
process (Ishwaran \&\ James, 2001\nocite{IshwaranJames01}). The mixing can
be done either on the intercept parameter, or on the entire vector of
regression coefficient parameters, depending on the user's choice. The
latter mixture model gives rise to a Dependent Dirichlet (DDP)\ process
mixture model (see DeIorio, et al. 2004\nocite%
{DeIorioMullerRosnerMacEachern04}). In principle any one of these DP\ or DDP
mixture models can also be used to perform causal inferences from a RDD\
design, using the inference methods discussed earlier in this chapter.

\bigskip 

\bigskip 

\noindent {\Large References}

\begin{description}
\item Aiken, L., West, S., Schwalm, D., Carroll, J., \& Hsiung, S. (1998).
Comparison of a randomized and two quasi-experimental designs in a single
outcome evaluation efficacy of a university-level remedial writing program. 
\textit{Evaluation Review}, \textit{22}, 207-244.

\item Angrist, J., \& Lavy, V. (1999). Using Maimonides'rule to estimate the
effect of class size on scholastic achievement. \textit{The Quarterly
Journal of Economics}, \textit{114}, 533-575.

\item Angrist, J., \& Lavy, V. (2008). \textit{Replication data for: Using
Maimonides' rule to estimate the effect of class size on student achievement.%
} \ http://thedata.harvard.edu/dvn/dv/JAngrist/. (Accessed: 2014-06-05)

\item Berk, R., Barnes, G., Ahlman, L., \& Kurtz, E. (2010). When second
best is good enough: A comparison between a true experiment and a regression
discontinuity quasi-experiment. \textit{Journal of Experimental Criminology}%
, \textit{6}, 191-208.

\item Black, D., Galdo, J., \& Smith, J. (2005). \textit{Evaluating the
regression discontinuity design using experimental data.} (Unpublished
manuscript)

\item Bloom, H. (2012). Modern regression discontinuity analysis. \textit{%
Journal of Research on Educational Effectiveness}, \textit{5}, 43-82.

\item Buddelmeyer, H., \& Skoufias, E. (2004). \textit{An evaluation of the
performance of regression discontinuity design on PROGRESA.} World Bank
Publications.

\item Cook, T. (2008). Waiting for life to arrive: A history of the
regression-discontinuity design in psychology, statistics and economics. 
\textit{Journal of Econometrics}, \textit{142}, 636-654. 

\item Cook, T., \& Campbell, D. (1979). \textit{Quasi-experimentation:
Design and analysis issues for field settings}. Chicago: Rand McNally.

\item Dawid, A. (1979). Conditional independence in statistical theory. 
\textit{Journal of the Royal Statistical Society}, \textit{Series B}, 
\textit{41}, 1-31.

\item Dawid, A. (2000). Causal inference without counterfactuals. \textit{%
Journal of the American Statistical Association}, \textit{95}, 407-424.

\item Dawid, A. (2002). Influence diagrams for causal modelling and
inference. \textit{International Statistical Review}, \textit{70}, 161-189.

\item Dawid, A. (2010). Beware of the DAG! \textit{Journal of Machine
Learning Research-Proceedings Track}, \textit{6}, 59-86.

\item DeIorio, M., M\"{u}ller, P., Rosner, G., \& MacEachern, S. (2004). An
ANOVA model for dependent random measures. \textit{Journal of the American
Statistical Association}, \textit{99}, 205-215.

\item Gelfand, A., \& Mukhopadhyay, S. (1995). On nonparametric Bayesian
inference for the distribution of a random sample. \textit{Canadian Journal
of Statistics}, \textit{23}, 411-420.

\item Geyer, C. (2011). Introduction to MCMC. In S. Brooks, A. Gelman, G.
Jones, \& X. Meng (Eds.), \textit{Handbook of Markov Chain Monte Carlo} (p.
3-48). Boca Raton, FL: CRC.

\item Goldberger, A. (2008/1972). Selection bias in evaluating treatment e%
\textcurrency ects: Some formal illustrations. In D. Millimet, J. Smith, \&
E. Vytlacil (Eds.), \textit{Modelling and evaluating treatment effects in
economics }(p. 1-31). Amsterdam: JAI Press.

\item Hahn, J., Todd, P., \& Klaauw, W. V. der. (2001). Identification and
estimation of treatment effects with a regression-discontinuity design. 
\textit{Econometrica}, \textit{69}, 201-209.

\item Hyamson, M. (1937). \textit{Annotated English translation of
Maimonides Mishneh Torah, Book I (The Book of Knowledge)}. New York: Jewish
Theological Seminary.

\item Imbens, G. (2004). Nonparametric estimation of average treatment e%
\textcurrency ects under exogeneity: A review. The Review of Economics and
Statistics, 86, 4-29.

\item Imbens, G.W., \& Lemieux, T. (2008). Regression discontinuity designs:
A guide to practice. \textit{Journal of Econometrics}, \textit{142}, 615-635.

\item Ishwaran, H., \& James, L. (2001). Gibbs sampling methods for
stick-breaking priors. \textit{Journal of the American Statistical
Association}, \textit{96}, 161-173.

\item Karabatsos, G. (2014a). \textit{Bayesian Regression: Nonparametric and
parametric models, version 2014x}. University of Illinois-Chicago.\newline
http://www.uic.edu/ georgek/HomePage/BayesSoftware.html.

\item Karabatsos, G. (2014b). Bayesian Regression: Nonparametric and
parametric models, version 2014b. Software users manual. University of
Illinois-Chicago.\newline
http://www.uic.edu/ georgek/HomePage/BayesSoftware.html.

\item Karabatsos, G., \& Walker, S. (2012). Adaptive-modal Bayesian
nonparametric regression. \textit{Electronic Journal of Statistics}, \textit{%
6}, 2038-2068.

\item Klaauw, W. V. der. (2008). Regression-discontinuity analysis: A survey
of recent developments in economics. \textit{Labour}, \textit{22}, 219-245.

\item Lauritzen, S. (2000). Causal inference from graphical models. In O.
Barndorff-Nielsen, D. Cox, \& C. Kluppelberg (Eds.), \textit{Complex
stochastic systems} (p. 63-107). London: CRC Press.

\item Lee, D. (2008). Randomized experiments from non-random selection in
U.S. house elections. \textit{Journal of Econometrics}, \textit{142},
675-697.

\item Lee, D., \& Lemieux, T. (2010). Regression discontinuity designs in
economics. \textit{The Journal of Economic Literature}, \textit{48}, 281-355.

\item LMT. (2012). \textit{Learning mathematics for teaching (LMT) assessment%
}. Ann Arbor, MI: University of Michigan.

\item Rosenbaum, P., \& Rubin, D. (1983). Assessing sensitivity to an
unobserved binary covariate in an observational study with binary outcome. 
\textit{Journal of the Royal Statistical Society}, \textit{Series B}, 
\textit{45}, 212-218.

\item Rubin, D. (1974). Estimating causal e\textcurrency ects of treatments
in randomized and nonrandomized studies. \textit{Journal of Educational
Psychology}, \textit{66}, 688-701.

\item Rubin, D. (1978). Bayesian inference for causal effects: The role of
randomization. \textit{Annals of Statistics}, \textit{6}, 34-58.

\item Rubin, D. (1990). Neyman (1923) and causal inference in experiments
and observational studies. \textit{Statistical Science}, \textit{5}, 472-480.

\item Rubin, D. (2008). For objective causal inference, design trumps
analysis. \textit{The Annals of Applied Statistics}, \textit{2}, 808-840.

\item Schochet, P. (2009). Statistical power for regression discontinuity
designs in education evaluations. \textit{Journal of Educational and
Behavioral Statistics}, \textit{34}, 238-266.

\item Shadish, W., Galindo, R., Wong, V., Steiner, P., \& Cook, T. (2011). A
randomized experiment comparing random and cutoff-based assignment. \textit{%
Psychological Methods}, \textit{16}, 179.

\item Stuart, A., \& Ord, K. (1998). \textit{Kendall's advanced theory of
statistics: Volume 1: Distribution theory.} New York: Wiley.

\item Thistlewaite, D., \& Campbell, D. (1960). Regression-discontinuity
analysis: An alternative to the ex-post facto experiment. \textit{Journal of
Educational Psychology}, \textit{51}, 309-317.

\item Trochim, W. (1984). \textit{Research design for program evaluation:
The regression-discontinuity approach.} Newbury Park, CA: Sage.

\item Wilk, M., \& Gnanadesikan, R. (1968). Probability plotting methods for
the analysis of data. \textit{Biometrika}, \textit{55}, 1-17. 

\item Wong, V., Steiner, P., \& Cook, T. (2013). Analyzing
regression-discontinuity designs with multiple assignment variables: A
comparative study of four estimation methods. \textit{Journal of Educational
and Behavioral Statistics}, \textit{38}, 107-141.
\end{description}

\newpage 

\begin{center}
\textbf{Figure Captions}

\bigskip 
\end{center}

\textbf{Figure 1.} \ For the LMT\ data, the posterior predictive density
estimates of $Y$, under treatment ($T=1$, red), and under non-treatment ($T=0
$, blue).

\textbf{Figure 2.} \ For the Maimonides' data, the posterior predictive
density estimates of $Y$, under treatment ($T=1$, red), and under
non-treatment ($T=0$, blue).

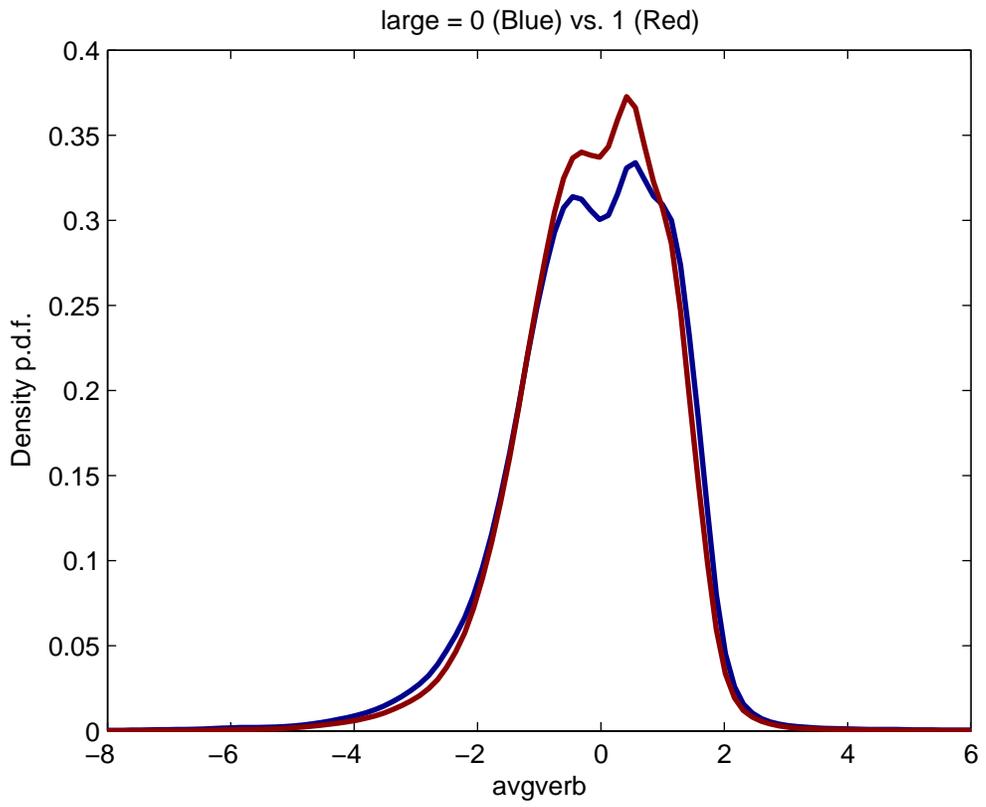
\end{document}